\documentclass[aps,prl,twocolumn,superscriptaddress,showpacs]{revtex4}
\usepackage{graphicx,graphics,amssymb,amsmath}
\usepackage{hyperref}

\begin{document}

\newcommand{\s}{\hspace{-1pt}}

\def\etal#1{, #1}
\def\tit#1{``#1,'' }
\def\jour#1{{\em #1}}

\title{Phonon-modulated magnetic interactions and spin Tomonaga-Luttinger liquid in the $p$-orbital antiferromagnet CsO$_2$}

\author{M. Klanj\v{s}ek}
\email{martin.klanjsek@ijs.si}
\affiliation{Jo\v{z}ef Stefan Institute, Jamova 39, 1000 Ljubljana, Slovenia}
\affiliation{EN-FIST Centre of Excellence, Trg OF 13, 1000 Ljubljana, Slovenia}

\author{D. Ar\v con}
\affiliation{Jo\v{z}ef Stefan Institute, Jamova 39, 1000 Ljubljana, Slovenia}
\affiliation{Faculty of Mathematics and Physics, University of Ljubljana, Jadranska 19, 1000 Ljubljana, Slovenia}

\author{A. Sans}
\affiliation{Max Planck Institute for Chemical Physics of Solids, N\"othnitzer Stra\ss e 40, 01187 Dresden, Germany}
\affiliation{Institute for Inorganic Chemistry, University of Stuttgart, Pfaffenwaldring 55, 70569 Stuttgart, Germany}

\author{P. Adler}
\affiliation{Max Planck Institute for Chemical Physics of Solids, N\"othnitzer Stra\ss e 40, 01187 Dresden, Germany}

\author{M. Jansen}
\affiliation{Max Planck Institute for Chemical Physics of Solids, N\"othnitzer Stra\ss e 40, 01187 Dresden, Germany}
\affiliation{Max Planck Institute for Solid State Research, Heisenbergstra\ss e 1, 70569 Stuttgart, Germany}

\author{C. Felser}
\affiliation{Max Planck Institute for Chemical Physics of Solids, N\"othnitzer Stra\ss e 40, 01187 Dresden, Germany}

\date{\today}



\begin{abstract}
The magnetic response of antiferromagnetic CsO$_2$, coming from the $p$-orbital $S=1/2$ spins of anionic O$_2^-$ molecules, is followed by $^{133}$Cs nuclear magnetic resonance across the structural phase transition occuring at $T_{s1}=61$~K on cooling. Above $T_{s1}$, where spins form a square magnetic lattice, we observe a huge, nonmonotonic temperature dependence of the exchange coupling originating from thermal librations of O$_2^-$ molecules. Below $T_{s1}$, where antiferromagnetic spin chains are formed as a result of $p$-orbital ordering, we observe a spin Tomonaga-Luttinger-liquid behavior of spin dynamics. These two interesting phenomena, which provide rare simple manifestations of the coupling between spin, lattice and orbital degrees of freedom, establish CsO$_2$ as a model system for molecular solids.
\end{abstract}

\maketitle

In many magnetic insulators, spins are well decoupled from other degrees of freedom, which implies simple Hamiltonians completely defined by the short-range magnetic exchange interactions. Model systems of this kind provide an excellent playground for the understanding of collective quantum phenomena, including
 Tomonaga-Luttinger liquid (TLL) in one-dimensional (1D) antiferromagnets~\cite{Giamarchi_2004}, Bose-Einstein condensation of magnons in dimer spin systems~\cite{Giamarchi_2008}, quantum criticality in gapped antiferromagnets~\cite{Ruegg2_2008,Mukhopadhyay_2012,Thede_2014,Kinross_2014} and spin-liquid behavior in frustrated spin systems~\cite{Balents_2010}.

\begin{figure*}
\includegraphics[width=1\linewidth]{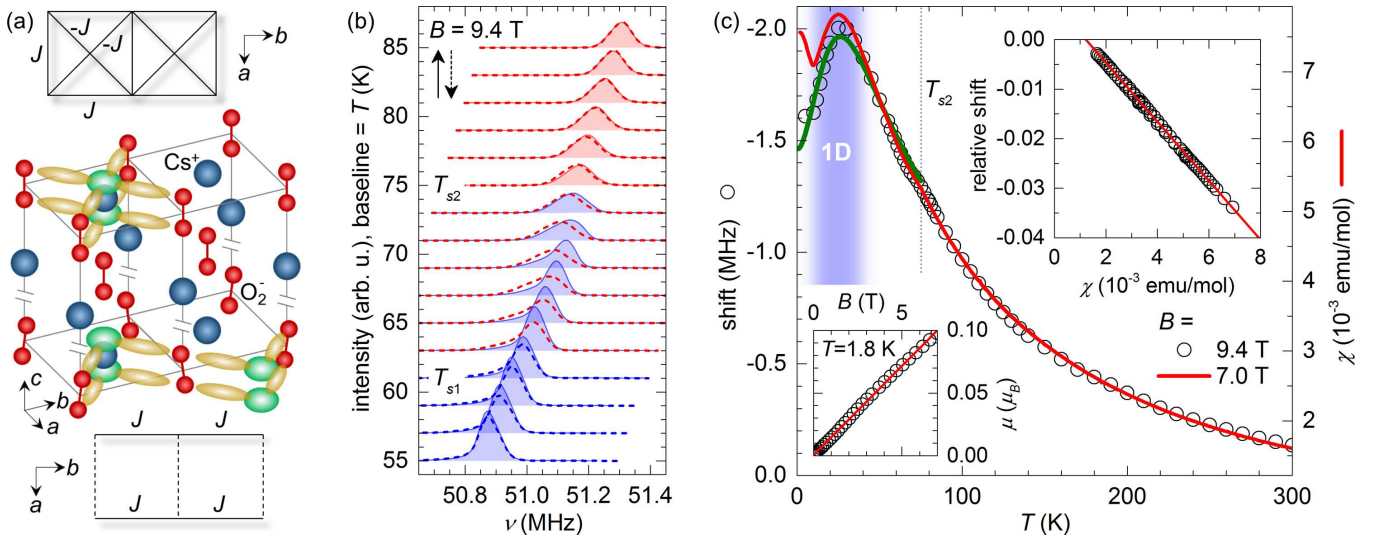}
\caption{(color online) Structural and magnetic properties of CsO$_2$. (a) Schematic crystal structure with representative O$_2^-$ $\pi_{x,y}^*$ orbitals (yellow) and Cs$^+$ $p_z$ orbitals (green). Above $T_{s2}$, the average direction of O$_2^-$ dumbbells is along the $c$ axis (top $ab$ layer) resulting in degenerate $\pi_{x,y}^*$ orbitals and frustrated-square magnetic lattice with exchange couplings $J$ ($-J$) between nearest (next-nearest) neighboring spins (top). Below $T_{s1}$, the tilt of O$_2^-$ dumbbells is staggered along the $b$ axis (bottom $ab$ layer) resulting in $\pi_{x,y}^*$ orbital ordering and formation of magnetic chains with exchange coupling $J$ along $b$ (bottom)~\cite{Riyadi_2012}. (b) Hysteretic evolution of $^{133}$Cs NMR spectrum across the structural phase transition taken on warming (solid lines, $T_{s2}=75$~K) and cooling (dashed lines, $T_{s1}=61$~K). The presence of high-$T$ (low-$T$) phase is indicated by red (blue) spectra. (c) $^{133}$Cs shift (open circles) follows static susceptibility $\chi$ (red line) down to $40$~K. Both are taken on warming. Shift in the low-$T$ phase is fitted with the chain model (green line) exhibiting a maximum as a signature of 1D spin correlations (blue background). Dotted vertical line indicates an anomaly at $T_{s2}$ due to the structural phase transition. Upper inset shows a linear dependence of the relative shift on $\chi$ above $40$~K, yielding a hyperfine coupling constant $A=-1.16$~T. Lower inset shows the field dependence of the O$_2^-$ magnetic moment $\mu$ calculated from the measured magnetization. Red line is a linear fit with $\partial\mu/\partial B=0.0142$~$\mu_B$/T.
}
\label{fig1}
\end{figure*}

In molecular solids, a class of magnetic insulators containing molecules as structural and magnetic units, spins cannot be decoupled from lattice and orbital degrees of freedom. This is particularly pronounced in systems based on small and light anionic O$_2^-$ molecules: alkali superoxides, $A$O$_2$ ($A$ = Na, K, Rb, Cs)~\cite{Zumsteg_1974,Labhart_1979,Bosch_1980,Riyadi_2012}, and alkali sesquioxides, $A_4$O$_6$ ($A$ = Rb, Cs)~\cite{Winterlik_2009_1,Winterlik_2009_2,Arcon_2013,Sans_2014}. Here, the O$_2^-$ anion carries an $S=1/2$ spin in a pair of $p$-derived degenerate $\pi^*$ orbitals~\cite{Solovyev_2008}. A strong coupling between spin, lattice and orbital degrees of freedom leads to complex physics~\cite{Winterlik_2009_1,Winterlik_2009_2,Arcon_2013,Sans_2014,Solovyev_2008,Nandy_2010,Ylvisaker_2010,Kim_2010,Wohlfeld_2011,Riyadi_2011,Kim_2014}, which is nevertheless based on two relatively simple mechanisms characteristic of molecular solids: (i)~the O$_2^-$ ``dumbbells'' can easily reorient, which modulates the overlaps of $\pi^*$ orbitals and thus the exchange coupling between the neighboring spins~\cite{Bosch_1980}; (ii)~the degeneracy of the $\pi^*$ orbitals is lifted by a structural phase transition involving the tilting of O$_2^-$ dumbbells, which is reminiscent of the Jahn-Teller effect~\cite{Solovyev_2008}. Calorimetric and magnetic studies indeed revealed several structural phase transitions in $A$O$_2$ systems back in the 1970s~\cite{Zumsteg_1974,Labhart_1979,Bosch_1980}, but their origin remained largely unexplained. These interesting observations were systematically revisited only in recent studies~\cite{Solovyev_2008,Nandy_2010,Ylvisaker_2010,Kim_2010,Wohlfeld_2011,Riyadi_2011,Kim_2014}. Among them, an important X-ray and Raman scattering study of CsO$_2$ clearly demonstrated the ordering of $p$ orbitals below the structural phase transition at $T_s\approx 70$~K, which leads to the formation of 1D antiferromagnetic spin-$1/2$ chains in an otherwise 2D magnetic lattice~\cite{Riyadi_2012}, as sketched in Fig.~\ref{fig1}(a). Two simple mechanisms mentioned above then lead to two obvious questions: (i)~given the high reorientational freedom of the O$_2^-$ dumbbells above $T_s$, does the associated librational phononic mode significantly affect the exchange coupling between the spins and (ii)~do the spin chains formed in the orbitally-ordered phase below $T_s$ exhibit a TLL behavior?

Concerning~(ii), TLL is a quantum-critical state predicted to be realized in gapless 1D spin systems. Its hallmark is the continuum of two-spinon excitations leading to the power-law behavior of various correlation functions~\cite{Giamarchi_2004}. The signatures of the TLL state were so far observed only in a few Cu- and Co-based 1D antiferromagnets~\cite{Lake_2005,Yoshida_2005,Ruegg1_2008,Klanjsek_2008,Thielemann1_2009,Hong_2010,Schmidiger_2012,Jeong_2013,Povarov_2015,Willenberg_2014,Kimura_2008,Klanjsek_2014}. In our nuclear magnetic resonance (NMR) experiment, we indeed observe a TLL behavior in the low-$T$, orbitally-ordered phase of CsO$_2$, which is a unique example of this exotic state in molecular solids. Concerning~(i), a dynamic modulation of the orbital overlaps by a phononic mode, a phenomenon still lacking a clear experimental demonstration, is predicted to lead to the temperature-dependent exchange coupling $J(T)$ with a {\em positive} slope as its fingerprint~\cite{Bramwell_1990}. In contrast, a static effect of lattice thermal expansion, also studied in the 1970s~\cite{Kennedy_1970,Zaspel_1977}, leads to $J(T)$ with a {\em negative} slope~\cite{Bramwell_1990}, simply because orbital overlaps become smaller as the lattice expands. The $J(T)$ between the neighboring spins in the high-$T$ phase of CsO$_2$ above $T_s$, as extracted from our NMR data, indeed exhibits a large positive slope, with a striking total increase of $50\%$, which we attribute to thermal librations of O$_2^-$ dumbbells. The effect is particularly pronounced as the involved magnetic molecules are small and thus highly reorientable.

The CsO$_2$ powder, prepared by oxidation of the freshly distilled Cs metal with dried molecular O$_2$ gas, was sealed in a glass tube for NMR and magnetization measurements. The sample was $\mathord{\sim}50\%$ enriched by $^{17}$O isotope for $^{17}$O NMR experiments. As these turned out to be difficult because of a very fast $^{17}$O spin-spin relaxation, we resorted to $^{133}$Cs NMR experiments. Fig.~\ref{fig1}(b) shows the temperature evolution of the $^{133}$Cs NMR spectrum in a magnetic field of $B=9.4$~T (with the Larmor frequency $52.461$~MHz) across the structural transition between the two phases. The transition, which is found to occur at $T_{s1}=61$~K on cooling and at $T_{s2}=75$~K on warming, is of first order, with hysteresis spanning the range of $\mathord{\sim}15$~K. The shift and width of the spectrum are related to the magnetic response of O$_2^-$ anions through the hyperfine coupling tensor ${\bf A}$. The shift is determined by the isotropic part of ${\bf A}$, while the width, found to be typically $\mathord{\sim}20$-times smaller than the shift, is determined by the correspondigly smaller anisotropic part of ${\bf A}$. A perfect linear relation between the shift and the magnetic susceptibility $\chi$ (taken in $B=7$~T) down to $40$~K [Fig.~\ref{fig1}(c) inset] yields the isotropic value $A=-1.16$~T for a single O$_2^-$ magnetic moment, using $g=2.1$ for the $g$-factor (an isotropic part of the measured $g$-tensor~\cite{Labhart_1979}) and assuming, for simplicity, the same coupling to all six neighboring O$_2^-$ moments [four in the $ab$ plane, two along $c$, see Fig.~\ref{fig1}(a)]. Below $40$~K, $\chi(T)$ ceases to follow the shift [Fig.~\ref{fig1}(c)], probably due to a small fraction of impurity spins picked by $\chi(T)$ as a bulk probe. A broad maximum in both datasets marks the low-$T$ onset of 1D spin correlations in an antiferromagnetic spin-$1/2$ chain~\cite{Eggert_1994}. The temperature dependence of the shift below $T_{s1}$ can indeed be perfectly fitted with the chain model~\cite{Feyerherm_2000} [Fig.~\ref{fig1}(c)], giving $J_{\rm 1D}/k_B=40.4$~K ($k_B$ is the Boltzmann constant)~\cite{Footnote} for the exchange coupling $J_{\rm 1D}$ between O$_2^-$ spins along the chain. At $1.8$~K, the lowest experimental temperature, the field-induced magnetic moment $\mu$ grows linearly with the magnetic field [lower inset of Fig.~\ref{fig1}(c)], as expected for the chain far from magnetic saturation~\cite{Giamarchi_2004}.

\begin{figure}
\includegraphics[width=1\linewidth]{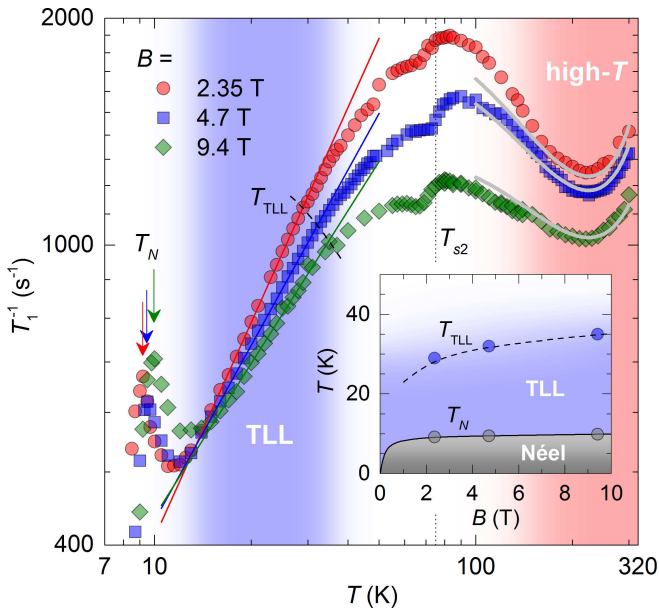}
\caption{(color online) Spin dynamics in CsO$_2$. $T_1^{-1}$ as a function of temperature $T$ taken on warming in three different magnetic fields $B$. Solid red, blue and green lines are power-law fits characteristic of the TLL behavior (blue background) valid in the range from $15$~K up to $T_{\rm TLL}$ (indicated by dashed line). Solid gray lines are the joint fit to the high-$T$ behavior (red background) for three magnetic field values. Arrows indicate the divergence in $T_1^{-1}(T)$ at $T_N$ due to the magnetic phase transition into the ordered N\'eel state. Dotted vertical line indicates the jump in $T_1^{-1}(T)$ at $T_{s2}=75$~K due to the structural phase transition. Inset outlines the low-$T$ phase diagram obtained from the data in the plot.}
\label{fig2}
\end{figure}

\begin{figure}
\includegraphics[width=1\linewidth]{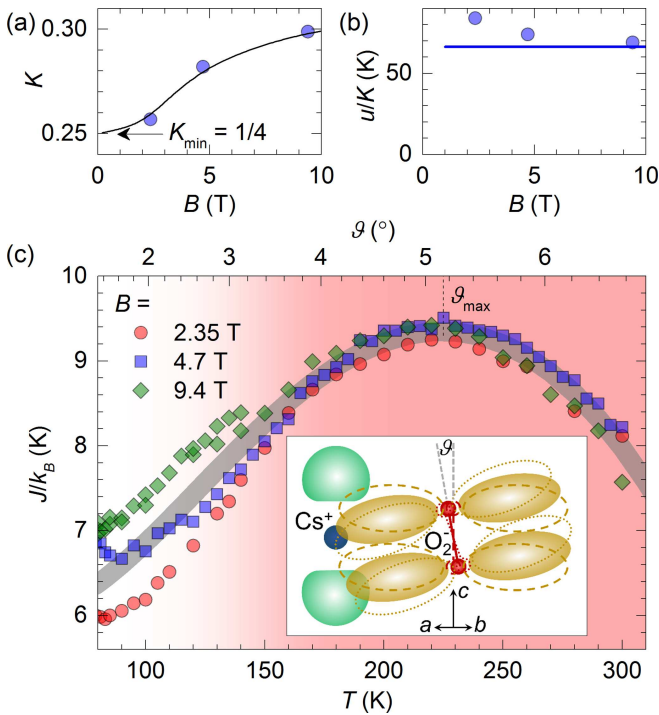}
\caption{(color online) Details of the TLL and high-$T$ behaviors extracted from Fig.~\ref{fig2}. Field dependence of (a) the TLL exponent $K$ (solid line is guide to the eye) and (b) the ratio $u/K$ compared to the prediction of Eq.~(\ref{eqTLL}) using the measured $\mu(B)$ [lower inset of Fig.~\ref{fig1}(b)] (solid line). (c) Temperature dependence of the exchange coupling $J(T)$ extracted from the data in Fig.~\ref{fig2} using Eq.~(\ref{eqT1para}). The best overlap of the three datasets is obtained for $\vert A\vert=0.82$~T. The upper horizontal scale is calculated using Eq.~(\ref{eqTheta}). Solid gray line is fit with the function $J(\vartheta)=J_0+J_2(\vartheta/\vartheta_0)^2+J_4(\vartheta/\vartheta_0)^4$ where $J_0/k_B=5.9$~K, $J_2/k_B=19$~K and $J_4/k_B=-28$~K. Inset shows a schematic dependence of the overlap between the O$_2^-$ $\pi_{x,y}^*$ orbital (yellow) and the Cs$^+$ $p_z$ orbital (green) on the tilt $\vartheta$ of the O$_2^-$ dumbbell from the $c$ axis. An optimal overlap (shaded yellow) is obtained for $\vartheta_{\rm max}=5.2^\circ$ reached at $225$~K where $J(T)$ exhibits a maximum.}
\label{fig3}
\end{figure}

To check whether the spin chains in the low-$T$ phase of CsO$_2$ exhibit a TLL behavior, we use NMR spin-lattice relaxation rate $T_1^{-1}$, which directly probes the low-frequency limit of the local spin-spin correlation function~\cite{Moriya_1956,Horvatic_2002}. As shown in Fig.~\ref{fig2}, $^{133}$Cs $T_1^{-1}(T)$ datasets measured in three different magnetic fields exhibit the power-law behavior characteristic of TLL up to the field-dependent temperature $T_{\rm TLL}$ of the order of $J_{\rm 1D}/k_B$. This behavior is outweighted below $\mathord{\sim}15$~K by the growth of 3D critical fluctuations preceding the 3D antiferromagnetic ordering~\cite{Labhart_1979,Riyadi_2012}. The transition occurs at the field-dependent N\'eel temperature $T_N$, which is marked by the characteristic peak in $T_1^{-1}(T)$. In the TLL state, transverse (i.e., perpendicular to the field) and longitudinal (i.e., parallel to the field) gapless spin fluctuations are possible~\cite{Okunishi_2007}. In CsO$_2$, the longitudinal fluctuations couple to $^{133}$Cs through the small anisotropic part of ${\bf A}$, so that their contribution to $T_1^{-1}$ is negligible with respect to the contribution of the transverse fluctuations coupled through the isotropic part $A$. In this case, the power-law dependence $T_1^{-1}=C(K) T^{1/(2K)-1}/u^{1/(2K)}$ is expected, where $K$ is the TLL exponent, $u$ is the velocity of spin excitations and $C(K)$ is the $K$-dependent prefactor~\cite{Klanjsek_2008,Giamarchi_1999,Supplemental}. The values of $K$ and $u$ are extracted as follows. First, the slope (in a log-log scale) of the $T_1^{-1}(T)$ datasets in Fig.~\ref{fig2} is given by $1/(2K)-1$, which directly defines the value of $K$. As $C(K)$ is then completely determined~\cite{Supplemental}, the value of $u$ follows directly from the vertical shift (in a log-log scale) of the $T_1^{-1}(T)$ datasets. We find $K$ to converge to $K_{\rm min}=1/4$ for $B=0$ [Fig.~\ref{fig3}(a)], in contrast to the value $1/2$ expected for the Heisenberg antiferromagnetic spin-$1/2$ chain~\cite{Giamarchi_2004}. The lowest possible value $K_{\rm min}$ is realized only in the presence of Ising-like exchange-coupling anisotropy~\cite{Giamarchi_2004,Okunishi_2007}. It is reached when the field is decreased to the critical field $B_c$, below which the chain dynamics becomes gapped. As the measured $\mu(B)$ in CsO$_2$ is linear down to $B\approx 0$ [lower inset of Fig.~\ref{fig1}(c)], $B_c$ should be very close to $0$, and the eventual exchange-coupling anisotropy should be small.

As the observed power-law behavior of $T_1^{-1}(T)$ in Fig.~\ref{fig2} covers less than a decade in temperature, the corresponding indication of the TLL behavior is only qualitative, and a quantitative check is needed. A stringent quantitative test is provided by the TLL-specific relation between the ratio $u/K$ derived from spin {\em dynamics} and the zero-$T$ susceptibility $\partial\mu/\partial B$ as a {\em static} observable~\cite{Giamarchi_2004}:
\begin{equation}\label{eqTLL}
	\frac{u}{K}=\frac{(g\mu_B)^2}{k_B}\frac{1}{\pi\frac{\partial\mu}{\partial B}}.
\end{equation}
Fig.~\ref{fig3}(b) shows a comparison between $u/K$ determined above from the $T_1^{-1}(T)$ datasets and the prediction of Eq.~(\ref{eqTLL}) using the field-independent value $\partial\mu/\partial B=0.0142$~T/$\mu_B$ extracted from the measured $\mu(B)$ [lower inset of Fig.~\ref{fig1}(c)]. The agreement is very good, although it gets slightly worse towards the critical field $B_c\approx 0$ where the TLL description is anyway expected to fail~\cite{Klanjsek_2008}. Furthermore, a field dependence of $T_{\rm TLL}$ and $T_N$ plotted in the inset of Fig.~\ref{fig2} reveals an expected phase diagram with the TLL behavior extending up to $T_{\rm TLL}<J_{\rm 1D}/k_B$ as in Ref.~\cite{Ruegg2_2008}. These results support the realization of the TLL state in the spin chains of CsO$_2$.

To extract $J(T)$ in the high-$T$ phase of CsO$_2$, we use $T_1^{-1}(T)$, which is expected to converge at high temperatures to a field- and temperature-independent value determined only by the exchange coupling~\cite{Moriya_1956}, a frequently observed behavior~\cite{Nath_2008,Nath_2009}. In contrast, our $T_1^{-1}(T)$ datasets in Fig.~\ref{fig2} exhibit an unusual, nonmonotonic and strongly field-dependent behavior above $T_{s2}$. The field dependence can be understood by realizing that the magnetic fields used in our experiment reach the energy scale comparable to the exchange coupling $J\approx J_{\rm 1D}/4$ expected in the high-$T$ phase (splitting the electron density between the two degenerate $p$ orbitals leads to the factor of $4$). In this case, the field-dependent Zeeman term for the electron spins cannot be neglected in comparison to the exchange term, as in the standard derivation of $T_1^{-1}(T,B)$~\cite{Moriya_1956}. By including both terms we obtain~\cite{Supplemental}
\begin{equation}\label{eqT1para}
	T_1^{-1}=\frac{\sqrt{\pi}}{2}\gamma^2\hbar A^2\frac{1}{\sqrt{zJ^2+(g\mu_B B)^2}},
\end{equation}
where $z$ is a number of neighboring O$_2^-$ spins to each O$_2^-$ spin, $\gamma/(2\pi)=5.585$~MHz/T is nuclear gyromagnetic ratio and $\hbar$ is reduced Planck constant. The exchange couplings along the sides of the square magnetic lattice in Fig.~\ref{fig1}(a) are antiferromagnetic~\cite{Riyadi_2012}, whereas the couplings along the diagonals are likely ferromagnetic due to the linear exchange path. For simplicity, we assume both to be of the same magnitude $J$ and so $z=8$. We also assume, as before, that $^{133}$Cs is coupled equally to six neighboring O$_2^-$ spins and add a factor of $6$ to the right side of Eq.~(\ref{eqT1para}). While Eq.~(\ref{eqT1para}) can explain the observed decrease of $T_1^{-1}$ with increasing $B$, the nonmonotonic behavior of $T_1^{-1}(T)$ can only be explained by postulating the temperature-dependent $J(T)$. If this postulate is correct, the three $J(T)$ datasets calculated from the $T_1^{-1}(T)$ datasets in Fig.~\ref{fig2} using Eq.~(\ref{eqT1para}) should overlap. Considering the temperature-independent~\cite{Supplemental} value of $A$ as a free parameter, the best overlap in the high-$T$ range above $150$~K [Fig.~\ref{fig3}(c)] is obtained for $\vert A\vert=0.82$~T, a value close to the one extracted in Fig.~\ref{fig1}(c). The corresponding joint $J(T)/k_B$ dataset, which is thus obtained directly from the experimental points, exhibits an unusual parabolical dependence with a maximum of $\mathord{\sim}9.3$~K, nicely matching the above estimated value $J_{\rm 1D}/(4k_B)=10.1$~K.

The obtained temperature dependence $J(T)$ cannot be of static origin, i.e., due to lattice thermal expansion. In such a case, the relative change of $J$ would be proportional to the relative change of the lattice parameter $a$ (or $b$), the coefficient being around $-12$~\cite{Bramwell_1990}. Based on the thermal expansion data for CsO$_2$~\cite{Riyadi_2012}, $J(T)$ would then exhibit a {\em monotonic negative} slope with the total drop of $\mathord{\sim}13\%$ between $100$~K and $300$~K. As this contribution is small compared to the observed {\em nonmonotonic} variation of around $50\%$, we neglect it. The obtained $J(T)$ is thus predominantly of dynamic origin, i.e., due to the modulation of the orbital overlaps by fast librations of O$_2^-$ dumbbells. A simple model treating the O$_2^-$ dumbbell as a harmonic oscillator supports this scenario. Denoting the O$_2^-$ dumbbell tilt from the $c$ axis as $\vartheta$, the quadratic mean $\sqrt{\left<\vartheta^2\right>}$ is calculated by writing the kinetic energy $\frac{1}{2}I\omega_l^2\left<\vartheta^2\right>$ as the thermal average over the oscillator states, $\hbar\omega_l/[e^{\hbar\omega_l/(k_BT)}-1]$, where $I=2.3\cdot 10^{-46}$~kgm$^2$ is the moment of inertia and $\omega_l=3.9\cdot 10^{13}$~s$^{-1}$ the libration frequency~\cite{Riyadi_2012}. This leads to
\begin{equation}\label{eqTheta}
	\sqrt{\left<\vartheta^2\right>}=\frac{\vartheta_0}{\sqrt{e^{T_0/T}-1}},
\end{equation}
where $\vartheta_0=\sqrt{2\hbar/(I\omega_l)}=8.7^\circ$ and $T_0=\hbar\omega_l/k_B=303$~K. Eq.~(\ref{eqTheta}) allows us to translate the observed $J(T)$ dependence into the $J(\vartheta)$ dependence (Fig.~\ref{fig3}, the upper horizontal scale) in the picture of frozen, tilted O$_2^-$ dumbbells (i.e., replacing $\sqrt{\left<\vartheta^2\right>}$ by $\vartheta$). Interestingly, the $J(\vartheta)$ dataset exhibits a maximum at $\vartheta_{\rm max}\approx 5.2^\circ$. In the proposed exchange path scenario over the Cs $p_z$ orbital~\cite{Riyadi_2012}, such a maximum is expected, as the optimal overlap between the O$_2^-$ $\pi_{x,y}^*$ orbital and the Cs$^+$ $p_z$ orbital is indeed achieved for a nonzero tilt angle [inset of Fig.~\ref{fig3}(c)]. Also the overall shape of the $J(\vartheta)$ dataset can be understood if we expand $J(\vartheta)=J_0+J_2(\vartheta/\vartheta_0)^2+J_4(\vartheta/\vartheta_0)^4$, where the odd terms are zero for symmetry reasons (i.e., tetragonal crystal symmetry~\cite{Riyadi_2012}) and we neglect higher-order terms. This $J(\vartheta)$ produces an excellent fit of the joint $J(\vartheta)$ dataset in Fig.~\ref{fig3}(c). Fig.~\ref{fig2} shows the corresponding joint fit of the three $T_1^{-1}$ datasets using Eqs.~(\ref{eqT1para}) and (\ref{eqTheta}). Finally, the obtained $\vartheta_{\rm max}$ value coincides with the static tilt of O$_2^-$ dumbbells in the low-$T$ phase as inferred from the X-ray and Raman scattering data~\cite{Riyadi_2012}. It thus appears that the low-$T$, orbitally-ordered phase maximizes the orbital overlaps and hence the exchange energy. As this appears to favor an exchange-driven origin of orbital ordering~\cite{Kugel_1982}, the issue should be investigated further.

In summary, we observed a TLL behavior of spin chains in the low-$T$, orbitally-ordered phase of CsO$_2$, which is a unique example of this exotic state in magnetic molecular solids. Moreover, in the high-$T$ phase, we observed a huge, nonmonotonic $J(T)$ dependence, which cannot be explained by static structural changes. Instead, it is consistent with the dynamic scenario based on O$_2^-$ molecular librations, which was suggested to be important also in solid O$_2$~\cite{Klotz_2010,Etters_1983}. Small molecular magnetic units in CsO$_2$ enable a particularly pronounced and clear demonstration of this phononic modulation of the exchange coupling. Nonetheless, a further neutron scattering study of CsO$_2$ is desired as it may lead to a better estimate of the TLL parameters~\cite{Lake_2005,Povarov_2015} and to an alternative check of the extracted $J(T)$. The observed two phenomena, which provide rare simple manifestations of the coupling between spin, lattice and orbital degrees of freedom, should help understand the behavior of a range of more complex molecular solids. To start with, magnetic susceptibility in the triangular-lattice organic molecular magnets implicitly suggests a positive slope of $J(T)$~\cite{Itou_2008,Isono_2014,Yoshida_2015}, likely originating in molecular librations.

\begin{acknowledgments}
We acknowledge the financial support by the European Union FP7-NMP-2011-EU-Japan project LEMSUPER under Contract No. 283214. We thank W.~Schnelle and R.~Koban for performing SQUID magnetization measurements.
\end{acknowledgments}

\newpage
\section{Supplemental material}

\subsection{NMR spin-lattice relaxation in the Tomonaga-Luttinger-liquid state}

{\em General expression.} To analyze the dynamics of the spin Tomonaga-Luttinger-liquid (TLL) state in CsO$_2$, we use an analytical expression for the NMR spin-lattice relaxation rate $T_1^{-1}$ derived for the case of dominant transverse spin fluctuations in the antiferromagnetic spin-$1/2$ chain~\cite{Klanjsek_2008}:
\begin{equation}\label{eqT1TLL}
	T_1^{-1}=\,\frac{\hbar\gamma^2 A^2 A_x}{k_B u}
		\cos\s\s\left(\s\frac{\pi}{4K}\s\right)
		B\s\s\left(\s\s \frac{1}{4K}, 1\s\s-\s\s\frac{1}{2K} \s\s\right)\s\s
		\left(\s\s \frac{2\pi T}{u} \s\right)^{\s\s\frac{1}{2K}-1},
\end{equation}
where $K$ is the interaction exponent, $u$ is the velocity of spin excitations (in kelvin units), $A_x$ is the amplitude of the transverse correlation function~\cite{Hikihara_2004}, $\gamma/(2\pi)$ is nuclear gyromagnetic ratio ($5.585$~MHz/T for $^{133}$Cs), $A$ is a relevant element of the hyperfine coupling tensor, $k_B$ is Boltzmann constant, $\hbar$ is reduced Planck constant and $B(x,y)=\Gamma(x)\Gamma(y)/\Gamma(x+y)$ where $\Gamma$ is the gamma function. The parameters $K$, $u$ and $A_x$ describe the TLL behavior, while the parameter $A$ describes the NMR response. In case of the CsO$_2$ powder, we use an isotropic part of the hyperfine coupling tensor $A=-1.16$~T [extracted from the scaling of the NMR shift with magnetic susceptibility in Fig.~\ref{fig1}(c)]. We need to add an extra factor of $4=2\cdot 2$ to the right side of Eq.~(\ref{eqT1TLL}): the first factor of $2$ because each $^{133}$Cs is coupled to two neighboring O$_2^-$ spins along the $c$ axis [Fig.~\ref{fig1}(a)] (while the total coupling to four neighboring O$_2^-$ spins in the $ab$ plane is zero due to antiferromagnetic nature of spin fluctuations in the chains), and the second factor of $2$ because two correlation functions (in two directions perpendicular to the chain) describe the transverse spin fluctuations.

{\em Extraction of the TLL parameters.} Among the three parameters $K$, $u$ and $A_x$ describing the TLL behavior, the dependence of $A_x$ on the field-induced magnetic moment $\mu$ was calculated in Ref.~\cite{Hikihara_2004} for various values of the exchange-coupling anisotropy $\Delta$. In our magnetic field range, $\mu\lesssim 0.14\mu_B$ [lower inset of Fig.~\ref{fig1}(c)], while $\Delta\approx 1$ in CsO$_2$. Fig.~2(a) in Ref.~\cite{Hikihara_2004} shows that in this case $A_x$ is almost field-independent and amounts to $0.12$. We use this value and extract the values of the remaining TLL parameters $u$ and $K$ from the $T_1^{-1}(T)$ datasets in Fig.~\ref{fig2} as follows. We confine our analysis to the temperature range $1.5T_N<T<3T_N$, with the corresponding $T_N$ for each dataset (inset of Fig.~\ref{fig2}), where the $T_1^{-1}(T)$ datasets appear to exhibit the power-law behavior. First, using Eq.~(\ref{eqT1TLL}), the slope (in a log-log scale) of the $T_1^{-1}(T)$ datasets is given by $1/(2K)-1$, which directly leads to the value of $K$. If we cast Eq.~(\ref{eqT1TLL}) into the form $T_1^{-1}=C(K) T^{1/(2K)-1}/u^{1/(2K)}$, the known value of $K$ then completely determines the value of $C(K)$. Finally, the value of $u$ follows directly from the vertical shift (in a log-log scale) of $T_1^{-1}(T)$ datasets.

\begin{figure}
\includegraphics[width=1\linewidth]{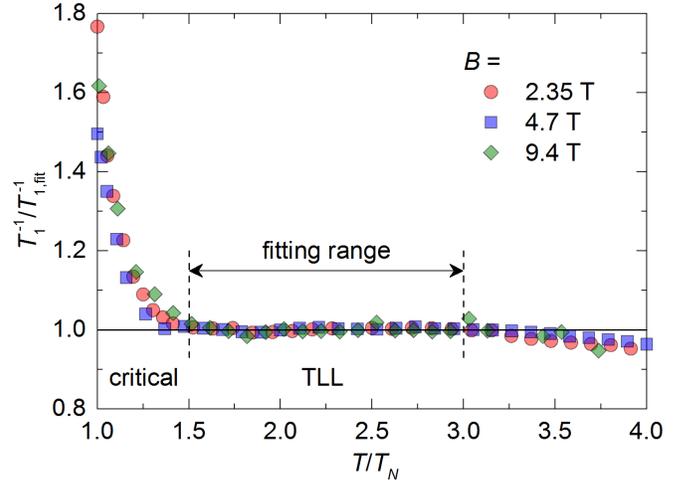}
\caption{$T_1^{-1}(T)$ datasets from Fig.~\ref{fig2} plotted with $T/T_N$ on a horizontal scale and $T_1^{-1}(T)/T_{1,{\rm fit}}^{-1}(T)$ on a vertical scale, where $T_{1,{\rm fit}}^{-1}(T)$ is given by Eq.~(\ref{eqT1TLL}) using the extracted values of $K$ and $u$.}
\label{figS1}
\end{figure}

{\em Demonstration of the power-law behavior.} To emphasize the quality of the corresponding power-law fits, the $T_1^{-1}(T)$ datasets are presented in a scaled form in Fig.~\ref{figS1}, i.e., $T_1^{-1}(T)/T_{1,{\rm fit}}^{-1}(T)$ as a function of $T/T_N$, where $T_{1,{\rm fit}}^{-1}(T)$ is given by Eq.~(\ref{eqT1TLL}) using the extracted values of $K$ and $u$. All three datasets collapse on an almost perfect flat line in the selected temperature range $1.5T_N<T<3T_N$, hence demonstrating the power-law behavior of the original $T_1^{-1}(T)$ datasets in this range. An identical approach to extract the field dependence of $K$ was recently used in Ref.~\cite{Jeong_2013}, where an analogous collapse on a flat line was obtained in a comparably broad temperature range $2T_N<T\lesssim 4T_N$. 

\subsection{NMR spin-lattice relaxation in the high-temperature paramagnetic state}

{\em Moriya's expression.} The calculation of the NMR spin-lattice relaxation rate $T_1^{-1}$ for the Heisenberg antiferromagnet in the paramagnetic state was carried out by Moriya in Ref.~\cite{Moriya_1956}. In this calculation, the fluctuations of the electron spins responsible for the NMR relaxation are derived for the field-independent exchange term only. We extend Moriya's calculation by including also the field-dependent Zeeman term for electron spins, which becomes relevant when the energy scales of the magnetic field $B$ and of the exchange coupling $J$ are comparable. This is the case in CsO$_2$ where $J/k_B$ is of the order of $10$~K and the magnetic fields used in the experiment are of the order of $B=10$~T, which translates to the value $g\mu_BB/k_B=14.1$~K ($\mu_B$ is Bohr magneton and $g=2.1$, an isotropic part of the measured $g$-tensor~\cite{Labhart_1979}) comparable to $J/k_B$. In a lot of materials studied so far, the exchange couplings are of the order of $100$~K or even $1000$~K, and in these cases the original Moriya's calculation has been applicable. In addition, we are interested in the high-temperature limit, i.e., $k_BT\gg J$. This limit is realized typically for $k_BT>10 J$, i.e., for $T>100$~K in CsO$_2$.

{\em Derivation of the extended Moriya's expression.} In case of an isotropic hyperfine coupling $A$ between the nuclear spin and the electron spin residing at site $l$, the NMR spin-lattice relaxation rate can be written~\cite{Horvatic_2002}
\begin{eqnarray}\label{eqT1iso}
	T_1^{-1} &=& \frac{1}{2}\gamma^2\int_{-\infty}^\infty {\rm d}t\,e^{i\omega_{\rm NMR}t}\cdot \nonumber\\
	&\cdot & \left\{ A^2\langle S_l^x(t)S_l^x\rangle + A^2\langle S_l^y(t)S_l^y\rangle \right\},
\end{eqnarray}
where $S_l^x$ and $S_l^y$ are spin operators, $t$ is time and $\omega_{\rm NMR}=\gamma B$ is the NMR frequency. We start by calculating the thermal spin correlation function $\langle S_l^x(t)S_l^x\rangle$. The time evolution of the spin operator is $S_l^x(t)=e^{iHt/\hbar}S_l^xe^{-iHt/\hbar}$ where
\begin{equation}
	H=\sum_j J_j\vec{S}_j\vec{S}_l-g\mu_BBS_l^z
\end{equation}
is a part of the hamiltonian relevant for the spin at site $l$, now containing both the Heisenberg and the Zeeman term. The sum runs over all neighboring spins. Following Moriya, each exponential factor is developed in a Taylor series over $t$ leading to
\begin{eqnarray}\label{eqS1}
	\langle S_l^x(t)S_l^x\rangle &=& \langle S_l^xS_l^x\rangle +
	i\frac{t}{\hbar}\langle [H,S_l^x]S_l^x\rangle + \nonumber\\
	&+&\frac{1}{2}\left(i\frac{t}{\hbar}\right)^2\langle [H,S_l^x][S_l^x,H]\rangle + \cdots
\end{eqnarray}
The thermal average for an arbitrary operator $P$ is calculated as $\langle P\rangle={\rm tr}\{e^{-\beta H}P\}$ where $\beta=1/(k_BT)$ and $T$ is temperature. In the high-temperature limit, we can approximate $e^{-\beta H}$ to $1$, so that $\langle P\rangle={\rm tr}\{P\}$. For the $S=1/2$ spin, the first term on the right side of Eq.~(\ref{eqS1}) then evaluates to $1/4$, the second term evaluates to zero, while in the third term ${\rm tr}\{[H,S_l^x][S_l^x,H]\}=\frac{1}{4}[(g\mu_BB)^2+\frac{1}{2}\sum_j J_j^2]$. When the exchange couplings to all of the $z$ neighboring electron spins are the same in magnitude, and this magnitude amounts to $J$, we can set $\sum_j J_j^2=zJ^2$. Finally, the correlation function can be written as
\begin{eqnarray}\label{eqCorr}
	\langle S_l^x(t)S_l^x\rangle &=& \frac{1}{4}
		\left[ 1-\frac{1}{4}\frac{t^2}{\hbar^2}\left\{zJ^2+2(g\mu_B B)^2\right\} \right]\approx \nonumber\\
	&\approx & \frac{1}{4}e^{-\frac{1}{4}\frac{t^2}{\hbar^2}\{zJ^2+2(g\mu_B B)^2\}},
\end{eqnarray}
where the first line has been recognized as the beginning of the Taylor expansion of the exponential function and the corresponding replacement has been made. This is the Gaussian approximation~\cite{Moriya_1956}. The result for the correlation function $\langle S_l^y(t)S_l^y\rangle$ is identical.

Capturing the result from Eq.~(\ref{eqCorr}) into a standard Gaussian form $\frac{1}{4}e^{-\omega_e^2t^2/2}$, we introduce the exchange frequency
\begin{equation}
	\omega_e=\frac{1}{\hbar}\sqrt{\frac{1}{2}zJ^2+(g\mu_B B)^2}.
\end{equation}
This is a characteristic frequency of the spin fluctuations. For $J/k_B\sim 10$~K and $B\sim 10$~T, $\omega_e$ is of the order of THz, much higher than $\omega_{\rm NMR}$ in the MHz range, meaning that we can set $\omega_{\rm NMR}=0$ in Eq.~(\ref{eqT1iso}). Plugging the result from Eq.~(\ref{eqCorr}) into Eq.~(\ref{eqT1iso}) and evaluating the integral, we obtain the final result
\begin{equation}\label{eqT1fin}
	T_1^{-1}=\frac{\sqrt{\pi}}{2}\gamma^2\hbar A^2\frac{1}{\sqrt{zJ^2+(g\mu_B B)^2}}.
\end{equation}
Setting $B=0$, this result recovers the well known and widely used Moriya's expression in the high-temperature approximation~\cite{Moriya_1956}. The expression found in textbooks is usually twice smaller as it is based on the correlation function for a single direction only, in contrast to both $x$ and $y$ directions considered here in Eq.~(\ref{eqT1iso}).

We note that $T_1^{-1}$ in Eq.~(\ref{eqT1fin}) exhibits a strong field dependence when the energy scales of $B$ and $J$ are comparable. The reason for this is not a modified $\omega_{\rm NMR}=\gamma B$, it is rather the change of spin dynamics. Namely, in presence of the sizeable magnetic field, the spin system becomes stiffer, the characteristic frequency $\omega_e$ of its fluctuations increases, meaning that the zero-frequency spectral density picked by $T_1^{-1}$ decreases. This is the physical meaning of Eq.~(\ref{eqT1fin}).

\begin{figure}
\includegraphics[width=1\linewidth]{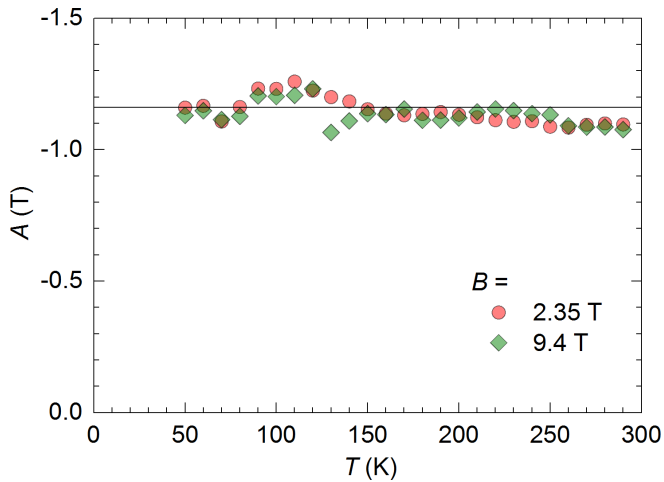}
\caption{Temperature dependence of the hyperfine coupling constant $A(T)$ for $B=9.4$~T (green diamonds) obtained from the data in Fig.~\ref{fig1}(c) inset using a sliding window centered at $T$, with the width of $30$~K ($80$~K) below (above) $150$~K. Similarly obtained $A(T)$ dataset for $B=2.35$~T (red circles) is plotted for comparison. Solid line represents the average value $A=-1.16$~T obtained in Fig.~\ref{fig1}(c).}
\label{figS2}
\end{figure}

{\em Temperature dependence of the hyperfine coupling constant.} As the exchange coupling $J$ between the neighboring O$_2^-$ spins is strongly temperature-dependent, a possibility should be considered that the same holds also for the hyperfine coupling constant $A$ between the $^{133}$Cs nucleus and the O$_2^-$ spin. Namely, the suggested exchange path between the O$_2^-$ spins goes through Cs$^+$~\cite{Riyadi_2012}. To check this possibility, we analyze in detail the linear dependence of the relative NMR shift on the susceptibility $\chi$ for $B=9.4$~T plotted in Fig.~\ref{fig1}(c) upper inset. The proportionallity constant between both variables is $A/(6N_A\mu_0)$, where $N_A$ is the Avogadro number, $\mu_0$ the vacuum permeability and $6$ accounts for the number of O$_2^-$ spins coupled to the $^{133}$Cs nucleus. We apply the linear fit in a sliding window centered at $T$, with the width of $30$~K below $150$~K, while above $150$~K, where the density of points is smaller, we use the width of $80$~K. The obtained value of $A$ is then assigned to the temperature $T$ defining the center of the window. The result displayed in Fig.~\ref{figS2} shows that the total variation of $A(T)$, amounting to $\mathord{\sim}10\%$, is negligible in comparison to the total variation of $J(T)$, so that $A$ can be considered temperature-independent. Due to a small fraction of impurity spins picked by $\chi(T)$ as a bulk probe and absent in the NMR shift as a local probe, $\chi(T)$ follows the NMR shift only down to $40$~K [Fig.~\ref{fig1}(c)]. Hence, our analysis could be carried out only down to $50$~K. No noticeable change of $A$ is observed upon the structural phase transition at $T_{s2}=75$~K (note that the data are taken on warming). For comparison, the same analysis is carried out for $B=2.35$~T, with the same result within the experimental error. It may seem surprising that the experimentally obtained hyperfine coupling constant $A$ is almost temperature-independent, while the $J(T)$ dependence is huge. This discrepancy likely comes from the fact that the hyperfine field is mediated by the Cs$^+$ $s$ orbital, while the exchange interaction is mediated by the Cs$^+$ $p_z$ orbital. The $\pi_{x,y}^*$(O$_2^-$)-$s$(Cs$^+$) orbital overlap apparently depends much less on the O$_2^-$ dumbbell tilt than the $\pi_{x,y}^*$(O$_2^-$)-$p_z$(Cs$^+$) orbital overlap.


\begin{thebibliography}{22}

\bibitem{Giamarchi_2004}
T.~Giamarchi, {\it Quantum Physics in One Dimension} (Oxford Univ. Press, Oxford, 2004).

\bibitem{Giamarchi_2008}
T.~Giamarchi, C.~R\"uegg, and O.~Tchernyshyov, \tit{Bose-Einstein condensation in magnetic insulators}\jour{Nature Phys.} {\bf 4}, 198 (2008).

\bibitem{Ruegg2_2008}
Ch.~R\"uegg\etal{B.~Normand, M.~Matsumoto, A.~Furrer, D.~F.~McMorrow, K.~W.~Kr\"amer, H.-U.~G\"udel, S.~N.~Gvasaliya, H.~Mutka, and M.~Boehm}, \tit{Quantum Magnets under Pressure: Controlling Elementary Excitations in TlCuCl$_3$}\jour{Phys. Rev. Lett.} {\bf 100}, 205701 (2008).

\bibitem{Mukhopadhyay_2012}
S.~Mukhopadhyay\etal{M.~Klanj\v sek, M.~S.~Grbi\'c, R.~Blinder, H.~Mayaffre, C.~Berthier, M.~Horvati\'c, M.~A.~Continentino, A.~Paduan-Filho, B.~Chiari, and O.~Piovesana}, \tit{Quantum-Critical Spin Dynamics in Quasi-One-Dimensional Antiferromagnets}\jour{Phys. Rev. Lett.} {\bf 109}, 177206 (2012).

\bibitem{Thede_2014}
M.~Thede\etal{A.~Mannig, M.~M\o ansson, D.~H\"uvonen, R.~Khasanov, E.~Morenzoni, and A.~Zheludev}, \tit{Pressure-Induced Quantum Critical and Multicritical Points in a Frustrated Spin Liquid}\jour{Phys. Rev. Lett.} {\bf 112}, 087204 (2014).

\bibitem{Kinross_2014}
A.~W.~Kinross\etal{M.~Fu, T.~J.~Munsie, H.~A.~Dabkowska, G.~M.~Luke, S.~Sachdev, and T.~Imai}, \tit{Evolution of Quantum Fluctuations Near the Quantum Critical Point of the Transverse Field Ising Chain System CoNb$_2$O$_6$}\jour{Phys. Rev. X} {\bf 4}, 031008 (2014).

\bibitem{Balents_2010}
L.~Balents, \tit{Spin liquids in frustrated magnets}\jour{Nature} {\bf 464}, 199 (2010).

\bibitem{Riyadi_2012}
S.~Riyadi\etal{B.~Zhang,~R.~A.~de~Groot, A.~Caretta, P.~H.~M.~van~Loosdrecht, T.~T.~M.~Palstra, and G.~R.~Blake}, \tit{Antiferromagnetic $S=1/2$ Spin Chain Driven by $p$-Orbital Ordering in CsO$_2$}\jour{Phys. Rev. Lett.} {\bf 108}, 217206 (2012).

\bibitem{Zumsteg_1974}
A.~Zumsteg, \tit{Magnetische und kalorische Eigenschaften von Alkali-Hyperoxid-Kristallen}\jour{PhD thesis} (ETH Z\"urich, 1973).

\bibitem{Labhart_1979}
M.~Labhart\etal{D.~Raoux, W.~K\"anzig, and M.~A.~B\"osch}, \tit{Magnetic order in $2p$-electron systems: Electron paramagnetic resonance and antiferromagnetic resonance in the alkali hyperoxides KO$_2$, RbO$_2$, and CsO$_2$}\jour{Phys. Rev. B} {\bf 20}, 53 (1979).

\bibitem{Bosch_1980}
M.~A.~B\"osch, M.~E.~Lines, and M.~Labhart, \tit{Magnetoelastic Interactions in Ionic $\pi$-Electron Systems}\jour{Phys. Rev. Lett.} {\bf 45}, 140 (1980).

\bibitem{Winterlik_2009_1}
J.~Winterlik\etal{G.~H.~Fecher, C.~A.~Jenkins, C.~Felser, C.~M\"uhle, K.~Doll, M.~Jansen, L.~M.~Sandratskii, and J.~K\"ubler}, \tit{Challenge of Magnetism in Strongly Correlated Open-Shell $2p$ Systems}\jour{Phys. Rev. Lett.} {\bf 102}, 016401 (2009).

\bibitem{Winterlik_2009_2}
J.~Winterlik\etal{G.~H.~Fecher, C.~A.~Jenkins, S.~Medvedev, C.~Felser, J.~K\"ubler, C.~M\"uhle, K.~Doll, M.~Jansen, T.~Palasyuk, I.~Trojan, M.~I.~Eremets, and F.~Emmerling}, \tit{Exotic magnetism in the alkali sesquioxides Rb$_4$O$_6$ and Cs$_4$O$_6$}\jour{Phys. Rev. B} {\bf 79}, 214410 (2009).

\bibitem{Arcon_2013}
D.~Ar\v con\etal{K.~Anderle, M.~Klanj\v sek, A.~Sans, C.~M\"uhle, P.~Adler, W.~Schnelle, M.~Jansen, and C.~Felser}, \tit{Influence of O$_2$ molecular orientation on $p$-orbital ordering and exchange pathways in Cs$_4$O$_6$}\jour{Phys. Rev. B} {\bf 88}, 224409 (2013).

\bibitem{Sans_2014}
A.~Sans\etal{J.~Nuss, G.~H.~Fecher, C.~M\"uhle, C.~Felser, and M.~Jansen}, \tit{Structural Implications of Spin, Charge, and Orbital Ordering in Rubidium Sesquioxide, Rb$_4$O$_6$}\jour{Z. anorg. allg. Chem.} {\bf 640}, 1239 (2014).

\bibitem{Solovyev_2008}
I.~V.~Solovyev, \tit{Spin-orbital superexchange physics emerging from interacting oxygen molecules in KO$_2$}\jour{New J. Phys.} {\bf 10}, 013035 (2008).

\bibitem{Nandy_2010}
A.~K.~Nandy\etal{P.~Mahadevan, P.~Sen, and D.~D.~Sarma},\tit{KO$_2$: Realization of Orbital Ordering in a $p$-Orbital System}\jour{Phys. Rev. Lett.} {\bf 105}, 056403 (2010).

\bibitem{Ylvisaker_2010}
E.~R.~Ylvisaker, R.~R.~P.~Singh, and W.~E.~Pickett, \tit{Orbital order, stacking defects, and spin fluctuations in the $p$-electron molecular solid RbO$_2$}\jour{Phys. Rev. B} {\bf 81}, 180405(R) (2010).

\bibitem{Kim_2010}
M.~Kim\etal{B.~H.~Kim, H.~C.~Choi, and B.~I.~Min}, \tit{Antiferromagnetic and structural transitions in the superoxide KO$_2$ from first principles: A $2p$-electron system with spin-orbital-lattice coupling}\jour{Phys. Rev. B} {\bf 81}, 100409(R) (2010).

\bibitem{Wohlfeld_2011}
K.~Wohlfeld, M.~Daghofer and A.~M.~Ole\'s, \tit{Spin-orbital physics for $p$ orbitals in alkali RO$_2$ hyperoxides--Generalization of the Goodenough-Kanamori rules}\jour{Europhys. lett.} {\bf 96}, 27001 (2011).

\bibitem{Riyadi_2011}
S.~Riyadi\etal{S.~Giriyapura, R.~A.~de~Groot, A.~Caretta, P.~H.~M.~van~Loosdrecht, T.~T.~M.~Palstra, and G.~R.~Blake}, \tit{Ferromagnetic Order from $p$-Electrons in Rubidium Oxide}\jour{Chem. Mater.} {\bf 23}, 1578 (2011).

\bibitem{Kim_2014}
M.~Kim and B.~I.~Min, \tit{Temperature-dependent orbital physics in a spin-orbital-lattice-coupled $2p$ electron Mott system: The case of KO$_2$}\jour{Phys. Rev. B} {\bf 89}, 121106(R) (2014).

\bibitem{Lake_2005}
B.~Lake\etal{D.~A.~Tennant, C.~D.Frost and S.~E.~Nagler}, \tit{Quantum criticality and universal scaling of a quantum antiferromagnet}\jour{Nature Mater.} {\bf 4}, 329 (2005).

\bibitem{Yoshida_2005}
Y.~Yoshida\etal{N.~Tateiwa, M.~Mito, T.~Kawae, K.~Takeda, Y.~Hosokoshi, and K.~Inoue}, \tit{Specific Heat Study of an $S=1/2$ Alternating Heisenberg Chain System: F$_5$PNN in a Magnetic Field}\jour{Phys. Rev. Lett.} {\bf 94}, 037203 (2005).

\bibitem{Ruegg1_2008}
Ch.~R\"uegg\etal{K.~Kiefer, B.~Thielemann, D.~F.~McMorrow, V.~Zapf, B.~Normand, M.~B.~Zvonarev, P.~Bouillot, C.~Kollath, T.~Giamarchi, S.~Capponi, D.~Poilblanc, D.~Biner, and K.~W.~Kr\"amer}, \tit{Thermodynamics of the Spin Luttinger Liquid in a Model Ladder Material}\jour{Phys. Rev. Lett.} {\bf 101}, 247202 (2008).

\bibitem{Klanjsek_2008}
M.~Klanj\v{s}ek\etal{H.~Mayaffre, C.~Berthier, M.~Horvati\'c, B.~Chiari, O.~Piovesana, P.~Bouillot, C.~Kollath, E.~Orignac, R.~Citro, and T.~Giamarchi}, \tit{Controlling Luttinger Liquid Physics in Spin Ladders under a Magnetic Field}\jour{Phys. Rev. Lett.} {\bf 101}, 137207 (2008).

\bibitem{Thielemann1_2009}
B.~Thielemann\etal{Ch.~R\"uegg, H.~M.~R\o nnow, A.~M.~L\"auchli, J.-S.~Caux, B.~Normand, D.~Biner, K.~W.~Kr\"amer, H.-U.~G\"udel, J.~Stahn, K.~Habicht, K.~Kiefer, M.~Boehm, D.~F.~McMorrow, and J.~Mesot}, \tit{Direct Observation of Magnon Fractionalization in the Quantum Spin Ladder}\jour{Phys. Rev. Lett.} {\bf 102}, 107204 (2009).

\bibitem{Hong_2010}
T.~Hong\etal{Y.~H.~Kim, C.~Hotta, Y.~Takano, G.~Tremelling, M.~M.~Turnbull, C.~P.~Landee, H.-J.~Kang, N.~B.~Christensen, K.~Lefmann, K.~P.~Schmidt, G.~S.~Uhrig, and C.~Broholm}, \tit{Field-Induced Tomonaga-Luttinger Liquid Phase of a Two-Leg Spin-$1/2$ Ladder with Strong Leg Interactions}\jour{Phys. Rev. Lett.} {\bf 105}, 137207 (2010).

\bibitem{Schmidiger_2012}
D.~Schmidiger\etal{P.~Bouillot, S.~M\"uhlbauer, S.~Gvasaliya, C.~Kollath, T.~Giamarchi, and A.~Zheludev}, \tit{Spectral and Thermodynamic Properties of a Strong-Leg Quantum Spin Ladder}\jour{Phys. Rev. Lett.} {\bf 108}, 167201 (2012).

\bibitem{Jeong_2013}
M.~Jeong\etal{H.~Mayaffre, C.~Berthier, D.~Schmidiger, A.~Zheludev, and M.~Horvati\'c}, \tit{Attractive Tomonaga-Luttinger Liquid in a Quantum Spin Ladder}\jour{Phys. Rev. Lett.} {\bf 111}, 106404 (2013).

\bibitem{Povarov_2015}
K.~Yu.~Povarov\etal{D.~Schmidiger, N.~Reynolds, R.~Bewley, and A.~Zheludev}, \tit{Scaling of temporal correlations in an attractive Tomonaga-Luttinger spin liquid}\jour{Phys. Rev. B} {\bf 91}, 020406(R) (2015).

\bibitem{Willenberg_2014}
B.~Willenberg\etal{H.~Ryll, K.~Kiefer, D.~A.~Tennant, F.~Groitl, K.~Rolfs, P.~Manuel, D.~Khalyavin, K.~C.~Rule, A.~U.~B.~Wolter, S.~S\"ullow}, \tit{Luttinger-Liquid Behavior in the Alternating Spin-Chain System Copper Nitrate}\jour{arXiv:} 1406.6149 (2014).

\bibitem{Kimura_2008}
S.~Kimura\etal{M.~Matsuda, T.~Masuda, S.~Hondo, K.~Kaneko, N.~Metoki, M.~Hagiwara, T.~Takeuchi, K.~Okunishi, Z.~He, K.~Kindo, T.~Taniyama, and M.~Itoh}, \tit{Longitudinal Spin Density Wave Order in a Quasi-1D Ising-Like Quantum Antiferromagnet}\jour{Phys. Rev. Lett.} {\bf 101}, 207201 (2008).

\bibitem{Klanjsek_2014}
M.~Klanj\v{s}ek\etal{M.~Horvati\'c, S.~Kr\"amer, S.~Mukhopadhyay, H.~Mayaffre, C.~Berthier, E.~Canevet, B.~Grenier, P.~Lejay, and E.~Orignac}, \tit{Giant magnetic-field dependence of the coupling between spin Tomonaga-Luttinger liquids in BaCo$_2$V$_2$O$_8$}\jour{arXiv:} 1412.2411 (2014).

\bibitem{Bramwell_1990}
S.~T.~Bramwell, \tit{Temperature dependence of the isotropic exchange constant}\jour{J. Phys.: Condens. Matter} {\bf 2}, 7527 (1990).

\bibitem{Kennedy_1970}
T.~A.~Kennedy, S.~H.~Choh, and G.~Seidel, \tit{Temperature Dependence of the Exchange Interaction in K$_2$CuCl$_4\cdot 2$H$_2$O}\jour{Phys. Rev. B} {\bf 2}, 3645 (1970).

\bibitem{Zaspel_1977}
C.~E.~Zaspel and J.~E.~Drumheller, \tit{Temperature dependence of the exchange interaction and applications to electron paramagnetic resonance}\jour{Phys. Rev. B} {\bf 16}, 1771 (1977).

\bibitem{Eggert_1994}
S.~Eggert, I.~Affleck, and M.~Takahashi, \tit{Susceptibility of the spin $1/2$ Heisenberg antiferromagnetic chain}\jour{Phys. Rev. Lett.} {\bf 73}, 332 (1994).

\bibitem{Feyerherm_2000}
R.~Feyerherm\etal{S.~Abens, D.~G\"unther, T.~Ishida, M.~Mei\ss ner, M.~Meschke, T.~Nogami, and M.~Steiner}, \tit{Magnetic-field induced gap and staggered susceptibility in the $S=1/2$ chain [PM$\cdot$Cu(NO$_3$)$_2\cdot$(H$_2$O)$_2$]$_n$ (PM = pyrimidine)}\jour{J. Phys.: Condens. Matter} {\bf 12}, 8495 (2000).

\bibitem{Footnote}
This is consistent with the determination $J_{1D}/2=20.0$~K in Ref.~\cite{Riyadi_2012} where the exchange coupling is defined as half of the prefactor to the exchange term.

\bibitem{Moriya_1956}
T.~Moriya, \tit{Nuclear Magnetic Relaxation in Antiferromagnetics}\jour{Prog. Theor. Phys.} {\bf 16}, 23 (1956).

\bibitem{Horvatic_2002}
M.~Horvati\'c and C.~Berthier, "NMR Studies of Low-Dimensional Quantum Antiferromagnets," in C.~Berthier, L.~P.~Levy, and G.~Martinez, {\it High Magnetic Fields: Applications in Condensed Matter Physics and Spectroscopy} (Springer-Verlag, Berlin, 2002), p. 200.

\bibitem{Giamarchi_1999}
T.~Giamarchi and A.~M.~Tsvelik, \tit{Coupled ladders in a magnetic field}\jour{Phys. Rev. B} {\bf 59}, 11398 (1999).

\bibitem{Okunishi_2007}
K.~Okunishi and T.~Suzuki, \tit{Field-induced incommensurate order for the quasi-one-dimensional $XXZ$ model in a magnetic field}\jour{Phys. Rev. B} {\bf 76}, 224411 (2007).

\bibitem{Supplemental}
See Supplemental Material for details of the $T_1^{-1}(T)$ analysis in the TLL and high-$T$ states.

\bibitem{Nath_2008}
R.~Nath\etal{A.~A.~Tsirlin, E.~E.~Kaul, M.~Baenitz, N.~B\"uttgen, C.~Geibel, and H.~Rosner}, \tit{Strong frustration due to competing ferromagnetic and antiferromagnetic interactions: Magnetic properties of $M$(VO)$_2$(PO$_4$)$_2$ ($M$ = Ca and Sr)}\jour{Phys. Rev. B} {\bf 78}, 024418 (2008).

\bibitem{Nath_2009}
R.~Nath\etal{Y.~Furukawa, F.~Borsa, E.~E.~Kaul, M.~Baenitz, C.~Geibel, and D.~C.~Johnston}, \tit{Single-crystal $^{31}$P NMR studies of the frustrated square-lattice compound Pb$_2$(VO)(PO$_4$)$_2$}\jour{Phys. Rev. B} {\bf 80}, 214430 (2009).

\bibitem{Hikihara_2004}
T.~Hikihara and A.~Furusaki, \tit{Correlation amplitudes for the spin-$1/2$ $XXZ$ chain in a magnetic field}\jour{Phys. Rev. B} {\bf 69}, 064427 (2004).

\bibitem{Kugel_1982}
K.~I.~Kugel' and D.~I.~Khomskii, \tit{The Jahn-Teller effect and magnetism: transition metal compounds}\jour{Sov. Phys. Usp.} {\bf 25}, 231 (1982).

\bibitem{Klotz_2010}
S.~Klotz\etal{Th.~Str\"assle, A.~L.~Cornelius, J.~Philippe, and Th.~Hansen}, \tit{Magnetic Ordering in Solid Oxygen up to Room Temperature}\jour{Phys. Rev. Lett.} {\bf 104}, 115501 (2010).

\bibitem{Etters_1983}
R.~D.~Etters, A.~A.~Helmy, and K.~Kobashi, \tit{Prediction of structures and magnetic orientations in solid $\alpha$- and $\beta$-O$_2$}\jour{Phys. Rev. B} {\bf 28}, 2166 (1983).

\bibitem{Itou_2008}
T.~Itou\etal{A.~Oyamada, S.~Maegawa, M.~Tamura, and R.~Kato}, \tit{Quantum spin liquid in the spin-$1/2$ triangular antiferromagnet EtMe$_3$Sb[Pd(dmit)$_2$]$_2$}\jour{Phys. Rev. B} {\bf 77}, 104413 (2008).

\bibitem{Isono_2014}
T.~Isono\etal{H.~Kamo, A.~Ueda, K.~Takahashi, M.~Kimata, H.~Tajima, S.~Tsuchiya, T.~Terashima, S.~Uji, and H.~Mori}, \tit{Gapless Quantum Spin Liquid in an Organic Spin-$1/2$ Triangular-Lattice $\kappa$-H$_3$(Cat-EDT-TTF)$_2$}\jour{Phys. Rev. Lett.} {\bf 112}, 177201 (2014).

\bibitem{Yoshida_2015}
Y.~Yoshida\etal{H.~Ito, M.~Maesato,	Y.~Shimizu, H.~Hayama, T.~Hiramatsu, Y.~Nakamura, H.~Kishida, T.~Koretsune, C.~Hotta, and G.~Saito}, \tit{Spin-disordered quantum phases in a quasi-one-dimensional triangular lattice}\jour{Nature Phys.}, advance online publication (2015).

\end{thebibliography}
\end{document}